\documentclass[a4paper,11pt]{article}

\usepackage{stmaryrd}
\usepackage{bm}
\usepackage{cite}
\usepackage{color}
\usepackage{dsfont}
\usepackage{nicefrac}
\usepackage{multirow}
\usepackage{graphics}
\usepackage{graphicx}
\usepackage{epsfig}
\usepackage{subfigure}
\usepackage{pgf}
\usepackage{amssymb}
\usepackage{amsmath,amssymb,amsfonts}
\usepackage{wasysym}

\begin{document}

\title{Inverse design method for metalens with optical vortices towards light focusing through localized phase retardation}

\author{Yongbo Deng$^1$\footnote{dengyb@ciomp.ac.cn}, Ulrike Wallrabe$^3$, Yihui Wu$^1$, Jan G. Korvink$^2$\footnote{jan.korvink@kit.edu}\\
1 State Key Laboratory of Applied Optics (SKLAO),\\
Changchun Institute of Optics, Fine Mechanics and Physics (CIOMP),\\
Chinese Academy of Sciences, Dongnanhu Road 3888, Changchun 130033, China\\
2 Institute of Microstructure Technology (IMT), \\
Karlsruhe Institute of Technology (KIT), \\
Hermann-von-Helmholtzplatz 1, Eggenstein-Leopoldshafen 76344, Germany\\
3 Laboratory for Microactuators, \\
Department of Microsystems Engineering (IMTEK), \\
University of Freiburg, Georges-K\"{o}hler-Allee 102, 79110 Freiburg, Germany}
\maketitle

\abstract{Metalenses can achieve diffraction-limited focusing through localized phase manipulation of the incoming light beam. Because these structures are ultrathin, less than a wavelength, this has the potential of achieving ultrathin optical elements, with a thickness limited mainly by the mechanical strength of the transparent substrate. Recently proposed metalenses are based on either dielectric nanofin arrays, or nanoparticles of large number, which leads to severe manufacturing challenges. To overcome these challenges, this paper predicts a new type of metalens with concentric-nanoring topology, where the number and size of the nanorings are determined using an inverse design method. By focusing the electrical field energy at a specified position, the convex-like metalens is inversely predicted with desired numerical aperture and a diffraction-limited focal spot. The Poynting vector distribution found demonstrates the mechanism of the lensing function, in which optical vortices are generated in the nanorings to achieve a matching of the phase and impedance between the substrate and free space, and further, to form a spherical wavefront and enhance the transmission of the optical energy. The inverse design method can also be extended to predict an axicon-like metalens with focal beam. The improved manufacturability is concluded from the geometry of the concentric-nanoring configurations.

\textbf{Keywords:} Convex-like metalens; axicon-like metalens; inverse design; concentric-nanoring topology; topology optimization}

\section{Introduction}\label{Introduction}

Ultrathin and flat metalenses, for diffraction-limited focusing of light, work in the visible spectrum and have been achieved based on metasurfaces \cite{Khorasaninejad2016}, which are composed of sub-wavelength-spaced phase shifters at an interface, and allowing for unprecedented control over the properties of light \cite{Yu2014,Kildishev2013}. Metalenses are less bulky than conventional lenses, and are more practical and less expensive to manufacture \cite{Yu2014}. Their features permit potential applications in microscopes, telescopes, cameras, smartphones, and other devices.

The essence of a lens is to pointwise modify the phase of an incident plane wave to form a focus. There are generally three methods to introduce phase compensation mechanisms into a lens design: shaping the surface of a piece of homogenous material, such as polishing a piece of glass to have a convex surface; use of diffractive structures to engineer the wavefront, as is done with a Fresnel zone plate \cite{Hecht1997}; using material inhomogeneities to modify the phase change in space, for instance, a gradient index (GRIN) lens \cite{Greegor2005}. A metalens applies similar phase compensation mechanisms to metamaterials to bring a plane-wave to a focus.

Optical metalenses have been designed based on arrays of optical antennas \cite{Yu2014}, arrays of nanoholes \cite{Huang2008}, optical masks \cite{Huang2009,Rogers2012,Fattal2010}, and nanoslits \cite{Verslegers2009}. Additionally, flat metamaterial-based hyperlenses and superlenses have been used to achieve sub-diffraction focusing \cite{Pendry2000,Smith2004,Liu2007,Cai2010}. These metalenses are designed based on the concept of an optical phase discontinuity, where the design of the metalens is obtained by imposing a spherical phase profile on the metasurface, and the control of wavefront is achieved via a phase shift experienced by the radiation as it scatters off the optically thin arrays of subwavelength-spaced resonators comprising the metasurface. As specified in \cite{Khorasaninejad2016}, these techniques do not provide the ability for complete control of the optical wavefront. Recently, a dielectric nanofin array-based metalens has been proposed in \cite{Khorasaninejad2016}, followed by a plasmonic nanoparticle-based metalens with a distribution of  nanoparticles determined by an evolutionary approach \cite{Hu2016}. These metalenses are composed of arrays of nanostructures (i.e. nanofins or nanoparticles) in large number, which potentially leads to manufacturing challenges.

This paper will show that it is possible to predict a new type of convex-like metalens using an inverse design method, which will achieve diffraction-limited focusing for the visible spectrum. Instead of imposing a phase profile on the metasurface, the inverse design method is used to find a geometrical configuration of the metalens with rotational symmetry, by maximizing the focused energy at the focal spot corresponding to a specified diameter and numerical aperture (NA) of the metalens. Topology optimization is utilized to implement an inverse design procedure. In the derived metalens, a titanium-dioxide material distribution with concentric-nanoring configuration focuses the plane wave by modifying the phase of the incident wave to form a spherical wavefront, which converges to the desired focal spot. In this viewpoint, an inverse design that maximizes the focused energy at the desired focal spot is equivalent to imposing a phase profile on the metasurface. By maximizing the localized energy deposition in a specified region, we show that the method can achieve compact axicon-like designs, with highly confined line-shaped foci.

\section{Metalens configured with concentric nanorings}\label{Metalens}

\begin{figure}[!htbp]
  \centering
  \includegraphics[width=0.7\columnwidth]{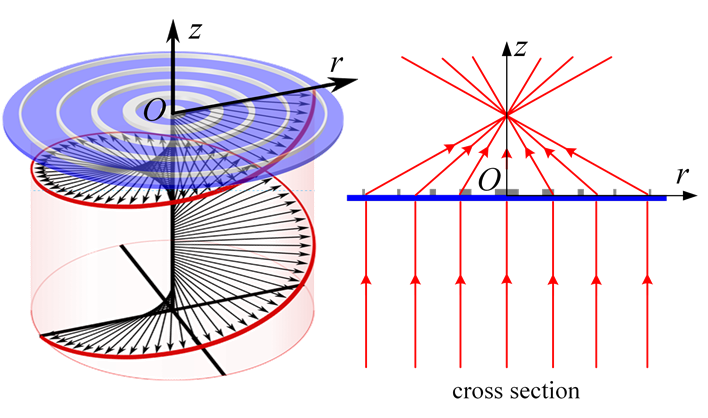}\\
  \caption{Sketch for the metalens with concentric-nanoring configuration, showing the coordinate system convention used in the paper. Linear or circularly polarised light is assumed to impinge vertically upon the optical element.} \label{Sketch}
\end{figure}

Metalenses that operate in the visible spectrum are achieved based on metasurfaces, which have feature sizes at the optical wavelength. The manufacturing processes most suitable and convienient for device construction are usually built upon nanolithography-like processes, e.g., electron-beam lithography (EBL), focused-ion beam (FIB) milling, interference lithography (IL), or nanoimprint lithography (NIL) \cite{Boltasseva2008}. These processes typically require the nanostructures to possess uniform cross section with sidewalls arranged perpendicular to the exposure or etching plane. For this reason, we confine the geometry to the concentric-nanoring configuration sketched in Figure \ref{Sketch}. The rotational symmetry of this geometrical configuration permits the decomposition of vertical illumination into components with circular and radial polarization. Subsequently, the optical field around a metalens can be reduced into a transverse wave propagating in the symmetry-plane of the metalens.

The substrate material of the metalens is a dielectric, specified to be glass in this paper. We select titanium dioxide ($\mathrm{TiO_2}$) for the ring material, because it is optically clear, interacts strongly with visible light, and has good manufacturability at the nanoscale.

By properly determining the number and size of the concentric nanorings of the metalens, the emerging secondary waves will constructively interfere and form a spherical wavefront which will arrive at the focal spot in phase, similarly to the waves that emerge from a conventional convex lens \cite{Hecht1997}. Conventional refractive optics employs a phase distribution created by a convex lens that focuses light to a single point in the limit of the paraxial (small angle) approximation. In contrast, the metalenses designed here produce a wavefront that remains spherical even for non-paraxial conditions. Therefore, we can achieve high NA focusing with negligible spherical aberrations.

Two features distinguish the proposed metalens from a Fresnel zone plate \cite{Hecht1997}. First, the wavefront shape for light focusing is accomplished within a distance comparable to the magnitude of the wavelength after the incident light traverses or is reflected by the optically thin metalens. Second, the metalens achieves its performance by matching its impedance with that of free space; thus optical vortices and transmission enhancement play a key role if the metalens material is a dielectric. In contrast, a Fresnel zone plate is designed with ring-shaped opaque zones and is based on diffraction theory, without considering subwavelength optical characteristics.

\section{Inverse prediction of designs}\label{InverseDesign}

In this section, we present the concepts for an inverse design method that will be used to determine the optimal number rings, and their width, for a specific metalens design.

\subsection{Modeling}\label{Modeling}
\begin{figure}[!htbp]
  \centering
  \includegraphics[width=0.5\columnwidth]{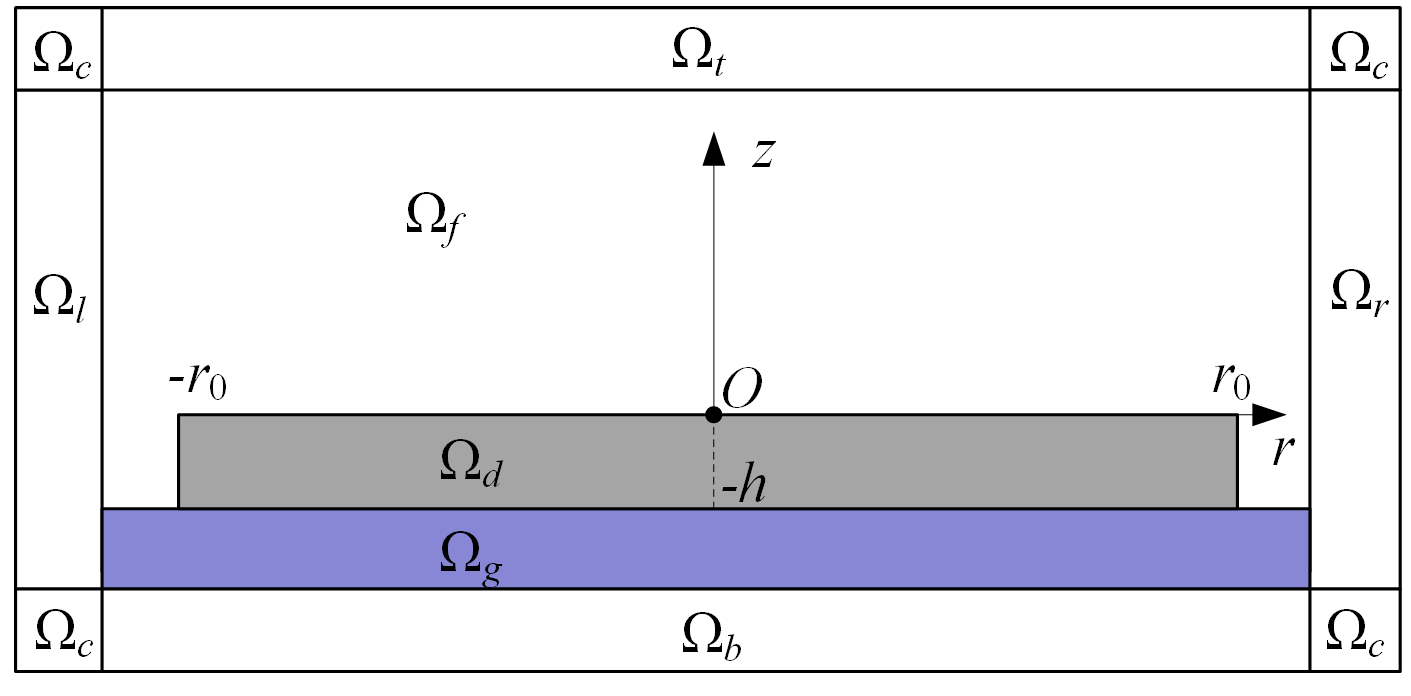}\\
  \caption{Sketch for the computational domain in the symmetry-plane of the metalens, in Cartesian coordinate system $rOz$. The infinite symmetry-plane is truncated by the PMLs $\Omega_l$, $\Omega_r$, $\Omega_t$, $\Omega_b$ and $\Omega_c$. The free space domain $\Omega_f$, the design domain $\Omega_d$, and substrate domain $\Omega_s$, are completely enclosed by the PMLs. The design domain $\Omega_d$ is the rectangle $\left[-r_0,r_0\right]\times\left[-h,0\right]$, where $r_0$ and $h$ are respectively the radius and thickness of the metalens.}\label{CrossSection}
\end{figure}

Because of the rotational symmetry of the concentric-nanoring configuration, and a decomposition of the light into circularly and radially polarized components (Figure \ref{Sketch}), the optical field in the symmetry-plane of the metalens can be described by the two-dimensional wave equation
\begin{equation}\label{WaveEquHz}
\begin{split}
    & \nabla\cdot \left[\epsilon_r^{-1}\nabla \left(H_s + H_i\right)\right] + k_0^2 \mu_r \left(H_s + H_i\right) = 0,~\mathrm{in}~\Omega
\end{split}
\end{equation}
where $H =H_s+H_i$ is the total field, $H_s$ is the scattered field; $H_i$ is the vertical incident wave on the interface between the glass substrate and the free space; $\epsilon_r$ and $\mu_r$ are the scalar relative permittivity and permeability, respectively; $k_0=\omega\sqrt{\epsilon_0\mu_0}$ is the free space wave number with $\omega$, $\epsilon_0$ and $\mu_0$ representing the angular frequency, free space permittivity and permeability, respectively; the time harmonic factor is $e^{j\omega t}$, with $j$ and $t$ respectively representing the imaginary unit and the time; $\Omega$ is the cross-sectional domain sketched in Figure \ref{CrossSection} in Cartesian coordinate system $rOz$, which includes free space denoted by $\Omega_f$, and the rectangular design domain $\Omega_d=\left[-r_0,r_0\right]\times\left[-h,0\right]$ for the cross-section of the metalens with radius and thickness respectively equal to $r_0$ and $h$, the substrate glass $\Omega_g$, and the surrounding perfectly matched layers (PMLs) used to truncate infinite space. The PMLs are implemented by a complex-valued coordinate transformation in the Cartesian coordinate system \cite{Jin2014}
\begin{equation}\label{CoordTransf}
\begin{split}
    \mathbf{p}' & = \left( \lambda\left(1-j\right) r/d, z \right),~\mathrm{in}~\Omega_l\cup\Omega_r \\
    \mathbf{p}' & = \left(r, \lambda\left(1-j\right) z/d \right),~\mathrm{in}~\Omega_t\cup\Omega_b \\
    \mathbf{p}' & = \left( \lambda\left(1-j\right) r/d, \lambda\left(1-j\right) z/d \right),~\mathrm{in}~\Omega_c \\
\end{split}
\end{equation}
where $\Omega_l$, $\Omega_r$, $\Omega_t$, $\Omega_b$ and $\Omega_c$ are sketched in Figure \ref{CrossSection}; $\mathbf{p}'=\left(r',z'\right)$ is the transformed coordinate; $\mathbf{p}=\left(r,z\right)$ is the original coordinate in the symmetry-plane of the metalens; $\lambda$ is the incident wavelength; $d$ is the thickness of the PMLs.

The inverse design procedure is implemented using a topology optimization approach. The metalens-material $\mathrm{TiO_2}$ is nonmagnetic. Hence, only the spatial distribution of relative permittivity needs to be determined in the metalens. In topology optimization, a structural configuration is represented by the distribution of a design variable $\rho\in\left[0,1\right]$, defined on the design domain. Here $\rho=1$ corresponds to the metalens-material, and $\rho=0$ to free space. Because of the quasi three dimensional nature of a concentric nanoring configuration, the design variable is defined on the design domain as a function of the $r$-coordinate only, and is considered uniform in the $z$ direction.

In our implementation of the inverse design method, the design variable is filtered by Helmholtz filters, which ensure smoothness of the material distribution, and numerical stability of the iterative procedure \cite{Lazarov2011,Wang2011}
\begin{equation}\label{HelmFilterGa}
\begin{split}
    -r_f^2 {\mathrm{d}^2\tilde{\rho} \over\mathrm{d} r^2} + \tilde{\rho} = & ~ \rho ,~\mathrm{in}~\Omega_d \\
    n_r {\mathrm{d} \tilde{\rho} \over\mathrm{d} r} = & ~ 0,~\mathrm{on}~\partial\Omega_d.
\end{split}
\end{equation}
Here, $r_f$ is the filter radius chosen based on numerical experiments \cite{Wang2011}; $\rho\in\left[0,1\right]$ is the design variable; $\tilde{\rho}$ is the filtered design variable; $n_r$ is the $r$-component of the unit outward normal on $\partial\Omega_d$. Furthermore, to ensure clear interfaces between the two materials of the derived structural topology, the filtered design variable is piecewise segmented \cite{Deng20161} and projected \cite{Guest2004} as follows
\begin{equation}\label{PiecewiseAndProjection}
\begin{split}
\bar{\rho}& = \sum_{n=1}^N {{\mathrm{tanh}\left(\beta\xi\right)+\mathrm{tanh}\left(\beta\left(\hat{\rho}_n-\xi\right)\right)}
     \over{\mathrm{tanh}\left(\beta\xi\right)+\mathrm{tanh}\left(\beta\left(1-\xi\right)\right)}},~\mathrm{with}~ \hat{\rho}_n =
\left\{
\begin{split}
& {1\over{V_{\Omega_n}}}\int_{\Omega_n} \tilde{\rho} \,\mathrm{d}v,~\mathrm{in}~\Omega_n,\\
& 0,~\mathrm{in}~\Omega_d\setminus\Omega_n.\\
\end{split}
\right.\\
\end{split}
\end{equation}
where $\mathrm{d}v=\mathrm{d}r\mathrm{d}z$ is the spatial differential. The design domain is divided into $N$ non-intersected pieces, satisfying $\Omega_d=\bigcup_{n=1}^N \Omega_n$; $\hat{\rho}$ is the segmented design variable; $\bar{\rho}$ is the projected design variable, which we name the physical density, taking the place of the design variable to represent the structural topology, and used to implement the material interpolation \cite{Wang2011}; $\xi\in\left[0,1\right]$ and $\beta$ are the threshold and projection parameters for the threshold projection. On the choice of values for $\xi$ and $\beta$, refer to \cite{Guest2004}. The physical density $\bar{\rho}$ is utilized to interpolate the relative permittivities between the metalens material and free space, corresponding to $\bar{\rho}=1$ and $\bar{\rho}=0$. The material interpolation is implemented as \cite{Deng2015}
\begin{equation}\label{Interpolation}
    \epsilon_r\left(\omega\right) = 10^{\log{\epsilon_{rm}\left(\omega\right)}-{{1-\bar{\rho}^3}\over{1+\bar{\rho}^3}}
    \left[\log{\epsilon_{rm}\left(\omega\right)}
    -\log{\epsilon_{rf}\left(\omega\right)}\right]},~\mathrm{in}~\Omega_d
\end{equation}
where $\epsilon_{rm}$ and $\epsilon_{rf}$ are the relative permittivities of $\mathrm{TiO_2}$ and free space, respectively.

In order to achieve the desired optical focusing performance, the inverse design objective is formulated to maximize the electric field density at the specified focal spot
\begin{equation}\label{FlatlensObj}
\begin{split}
J\left(\mathbf{E}\right) = \int_{\Omega_f} \left|\mathbf{E}\right|^2 \delta\left(\mathbf{p}-\mathbf{p}_f\right) \,\mathrm{d}v
\end{split}
\end{equation}
where $\mathbf{E}={1\over{j\omega}}\left(-{{\partial H}\over{\partial r}}\mathbf{e}_r+{{\partial H}\over{\partial z}}\mathbf{e}_z\right)$ is the electrical field; $\delta\left(\cdot\right)$ is the Dirac function; $\mathbf{p}_f\in\Omega_f$ is the position of the desired focal spot, which is determined based on the diameter of the metalens and the desired numerical aperture (NA). With these prerequisites, the inverse design problem is formulated as
\begin{equation}\label{TOOPProblem}
\begin{split}
& \mathrm{Find}~\rho~\mathrm{to}~\mathrm{maximize}~J\left(\mathbf{E}\right) \\
& \mathrm{Subject~to} \left\{
\begin{split}
& \nabla\cdot \left[\epsilon_r^{-1}\nabla \left(H_s + H_i\right)\right] + k_0^2 \mu_r \left(H_s + H_i\right) = 0,~\mathrm{in}~\Omega\\
& -r_f^2 {\mathrm{d}^2\tilde{\rho} \over\mathrm{d} r^2} + \tilde{\rho} = \rho ,~\mathrm{in}~\Omega_d\\
& 0 \leq \rho \leq 1,~\mathrm{in}~\Omega_d
\end{split}
\right.
\end{split}
\end{equation}
By solving the variational problem of equation \ref{TOOPProblem}, a geometrical configuration of a metalens can be derived, whose geometry corresponds to regions where $\bar{\rho}=1$.

\subsection{Analyzing and solving}\label{Analyzing}
Equation \ref{TOOPProblem} is solved using a gradient-based iterative procedure, where the gradient of the inverse design objective is used to iteratively evolve the design variable. The variational problem is analyzed using the adjoint method \cite{Hinze2009}, with the adjoint derivative of the electric field density at $\mathbf{p}_f$ derived to be
\begin{equation}\label{AdjointDerivative}
    \delta J = h \int_{-r_0}^{r_0} - \tilde{\rho}^a \delta\rho \,\mathrm{d}r
\end{equation}
where $\tilde{\rho}^a$ is the adjoint variable of $\tilde{\rho}$; $\delta\rho$ is the first order variational of $\rho$. The weak form adjoint equations of equation \ref{WaveEquHz} and \ref{HelmFilterGa} are solved for $\tilde{\rho}$
\begin{equation}\label{AdjEqus}
\begin{split}
&\mathrm{Find}~H_s^a~\mathrm{with}~\mathrm{Re}\left(H_s^a\right)\in\mathcal{H}\left(\Omega\right)~\mathrm{and}~ \mathrm{Im}\left(H_s^a\right)\in\mathcal{H}\left(\Omega\right),~\mathrm{and}~\tilde{\rho}^a\in\mathcal{H}\left(\Omega_d^r\right),~ \mathrm{satisfying:}\\
& \int_{\Omega_c} \delta \left(\mathbf{p} - \mathbf{p}_f\right) 2 \mathbf{E}^* \cdot {{\partial \mathbf{E}}\over{\partial \nabla H_s}} \cdot \nabla \phi + \epsilon_r^{-1} \nabla H_s^a \cdot \nabla \phi - k_0^2 \mu_r H_s^a \phi \,\mathrm{d}v \\
& + \int_{\Omega_{PMLs}} \epsilon_r^{-1} \left( {\partial\mathbf{p}'\over\partial\mathbf{p}} \nabla H_s^a \right) \cdot \left( {\partial\mathbf{p}'\over\partial\mathbf{p}} \nabla \phi \right) \left|{\partial\mathbf{p}'\over\partial\mathbf{p}}\right|^{-1} - k_0^2 \mu_r H_s^a \phi \left|{\partial\mathbf{p}'\over\partial\mathbf{p}}\right| \,\mathrm{d}v = 0,~\forall \phi \in \mathcal{H}\left(\Omega\right) \\
& \int_{-r_0}^{r_0} \left( r_f^2 {\mathrm{d} \tilde{\rho}^a \over \mathrm{d} r} {\mathrm{d} \psi \over \mathrm{d} r} + \tilde{\rho}^a \psi  + {\psi \over h}\int_{-h}^0\sum_{n=1}^N S_{\hat{\rho}_n}\,\mathrm{d}z \right) \,\mathrm{d}r = 0,~\forall\psi\in\mathcal{H}\left(\Omega_d^r\right)
\end{split}
\end{equation}
where $H_s^a$ is the adjoint variable of $H_s$; $\mathcal{H}\left(\Omega\right)$ and $\mathcal{H}\left(\Omega_d^r\right)$ are respectively the first order Hilbert space defined on $\Omega$ and $\Omega_d^r$; $\Omega_d^r=\left(-r_0,r_0\right)$ is the projection of $\Omega_d$ on the $r$-axis; $\Omega_{PMLs}$ is the union of the PMLs; $\mathrm{Re}\left(\cdot\right)$ and $\mathrm{Im}\left(\cdot\right)$ are the operators used to extract the real and imaginary parts of a complex variable; and $S_{\hat{\rho}_n}$ is defined to be
\begin{equation}
\begin{split}
    & S_{\hat{\rho}_n}\left(\Omega_n\right)  =  \left\{
    \begin{aligned}
    & {1 \over V_{\Omega_n}} \int_{\Omega_n} \mathrm{Re}\left({\partial {1\over\epsilon_r} \over \partial \bar{\rho}} {\partial \bar{\rho} \over \partial \hat{\rho}_n} \nabla H\right) \cdot \mathrm{Re}\left(\nabla H_s^a\right) \\
    & ~~~~~~ - \mathrm{Im}\left({\partial {1\over\epsilon_r} \over \partial \bar{\rho}} {\partial \bar{\rho} \over \partial \hat{\rho}_n} \nabla H \right) \cdot \mathrm{Im}\left(\nabla H_s^a\right) \,\mathrm{d}v,~\mathrm{in}~\Omega_n\\
    & 0,~\mathrm{in}~\Omega_d \setminus\Omega_n
    \end{aligned}\right.
\end{split}
\end{equation}
After adjoint analysis, the inverse design problem is solved iteratively using a numerical method. In our setup, the equations and corresponding adjoint equations are solved using the finite element method implemented in the commercial package COMSOL Multiphysics \cite{comsol}. The design variable is evolved using the method of moving asymptotes \cite{Svanberg1987}.

\section{Results and discussion}\label{Results}
\begin{figure}[!htbp]
  \centering
  \subfigure[$r_0=6\mu \mathrm{m}$]
  {\includegraphics[height=0.11\paperwidth]{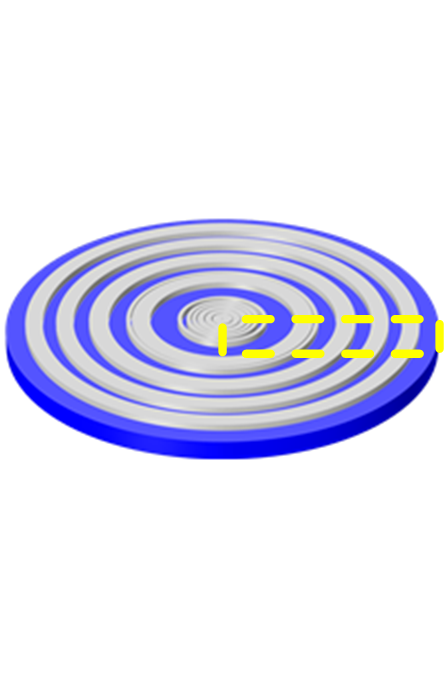}}
  \subfigure[$r_0=9\mu \mathrm{m}$]
  {\includegraphics[height=0.11\paperwidth]{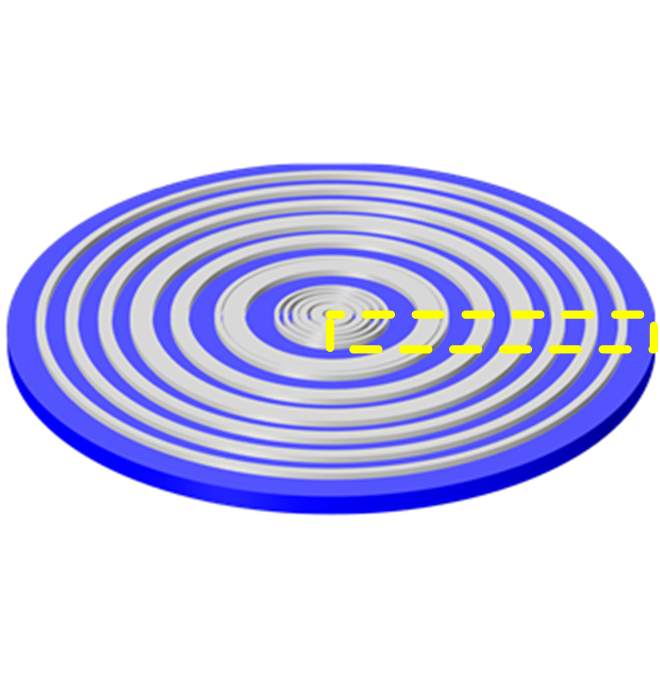}}
  \subfigure[$r_0=12\mu \mathrm{m}$]
  {\includegraphics[height=0.11\paperwidth]{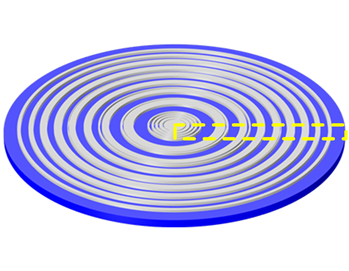}}
  \subfigure[$r_0=15\mu \mathrm{m}$]
  {\includegraphics[height=0.11\paperwidth]{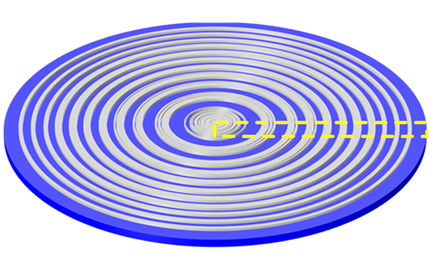}}
  \subfigure[$r_0=18\mu \mathrm{m}$]
  {\includegraphics[height=0.11\paperwidth]{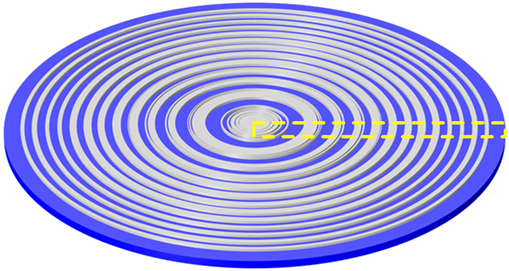}}\\
  \subfigure[Zoomed views of the regions marked by the dashed box in a$\sim$e.]
  {\includegraphics[height=0.25\paperwidth]{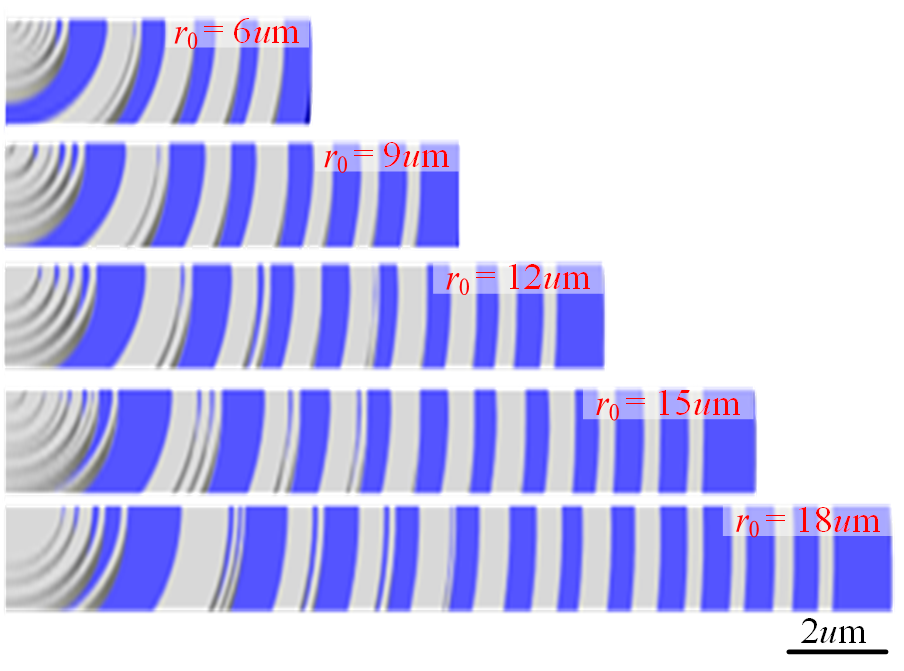}}
  \subfigure[Normalized electric field energy density in the focal plane.]
  {\includegraphics[height=0.28\paperwidth]{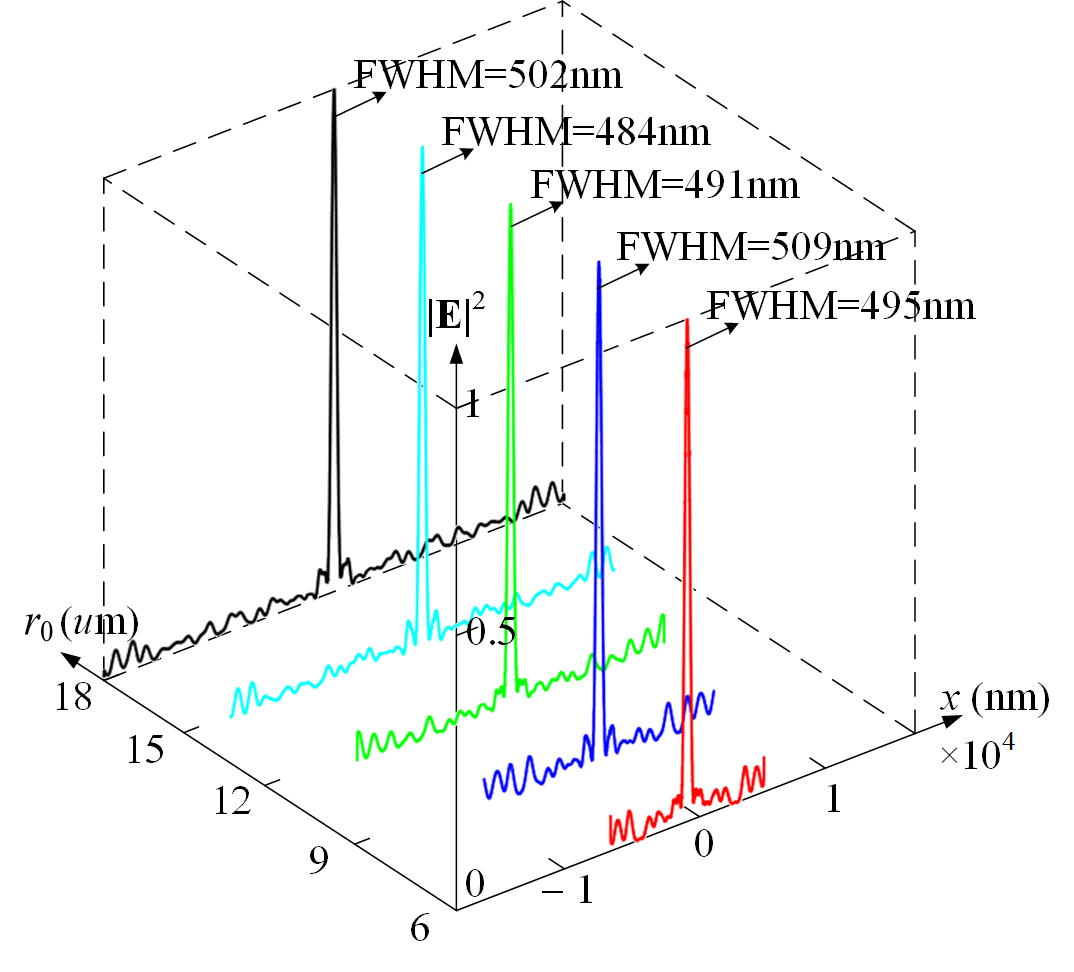}}
  \caption{Convex-like lenses with different ring radii, where the ring thickness is 200~nm, the numerical aperture is 0.8, and the incident wavelength is 700~nm.}\label{Lamb700H200D}
\end{figure}

\begin{figure}[!htbp]
  \centering
  \subfigure[NA=0.7]
  {\includegraphics[width=0.22\paperwidth]{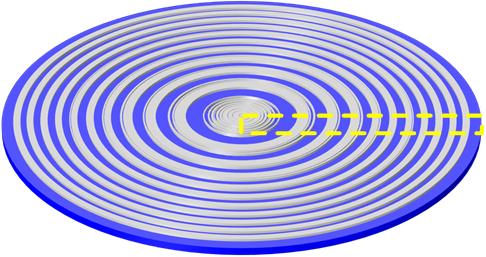}}
  \subfigure[NA=0.75]
  {\includegraphics[width=0.22\paperwidth]{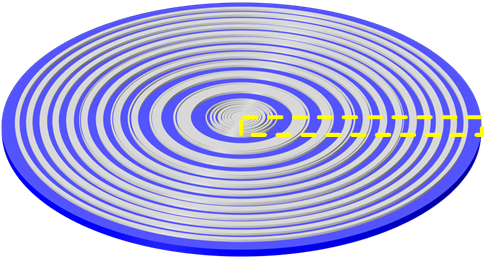}}
  \subfigure[NA=0.8]
  {\includegraphics[width=0.22\paperwidth]{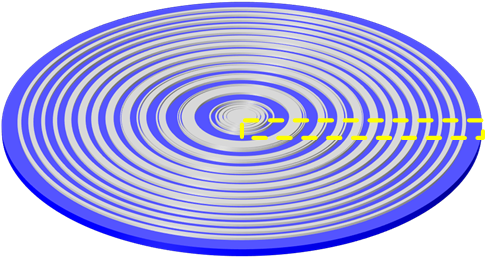}}
  \subfigure[NA=0.85]
  {\includegraphics[width=0.22\paperwidth]{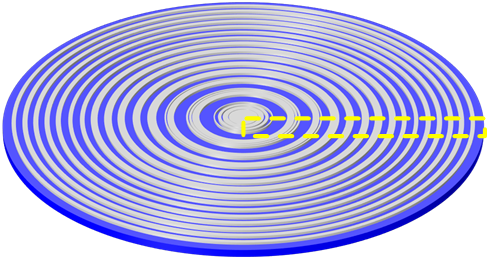}}
  \subfigure[NA=0.9]
  {\includegraphics[width=0.22\paperwidth]{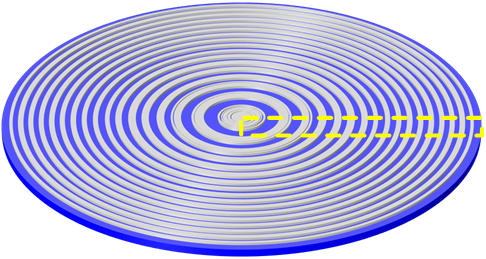}}\\
  \subfigure[Zoomed views of the regions marked by the dashed box in a$\sim$e.]
  {\includegraphics[height=0.25\paperwidth]{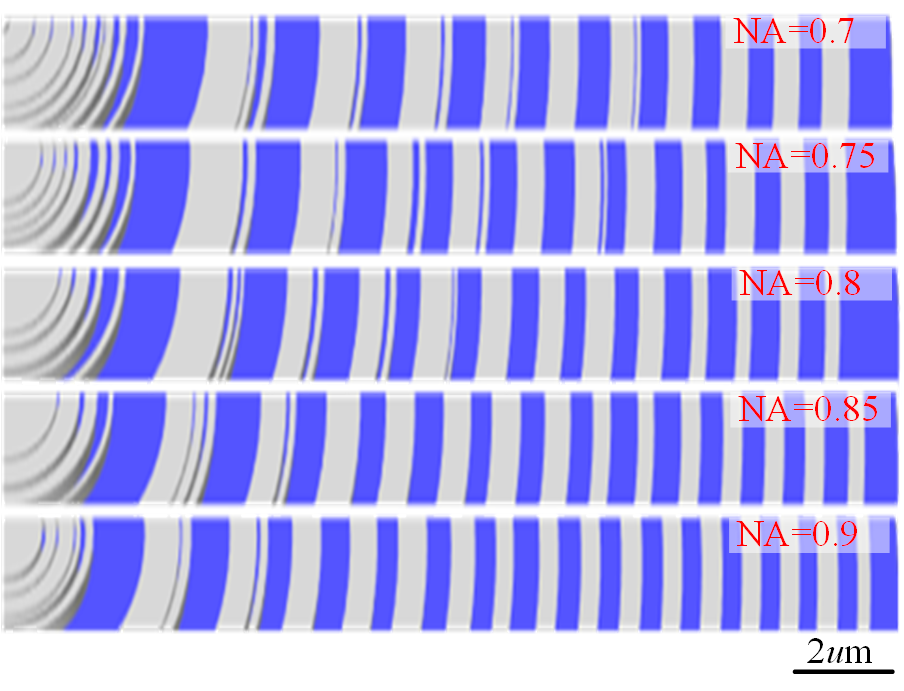}}
  \subfigure[Normalized electric field energy density in the focal plane.]
  {\includegraphics[height=0.28\paperwidth]{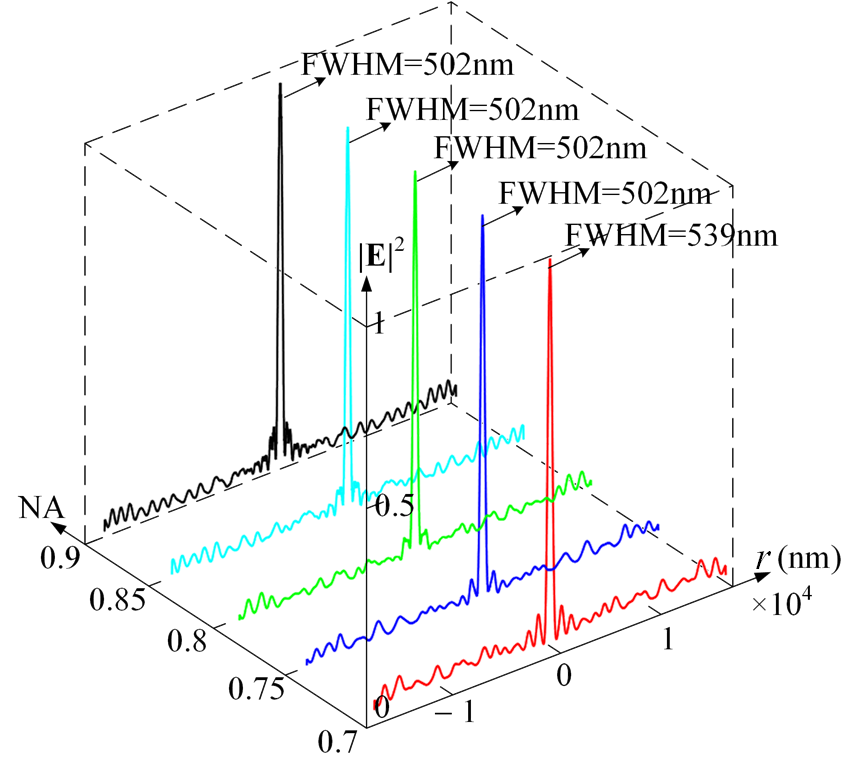}}
  \caption{Convex-like lenses with ring thickness 200~nm for an incident wavelength of 700~nm.}\label{FlatLenses200700}
\end{figure}

\begin{figure}[!htbp]
  \centering
  \subfigure[NA=0.7]
  {\includegraphics[width=0.22\paperwidth]{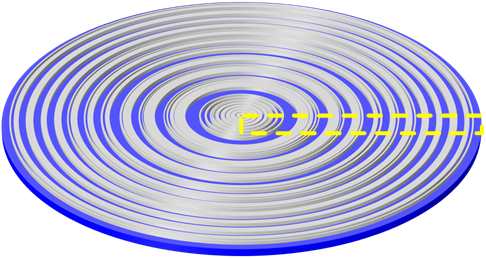}}
  \subfigure[NA=0.75]
  {\includegraphics[width=0.22\paperwidth]{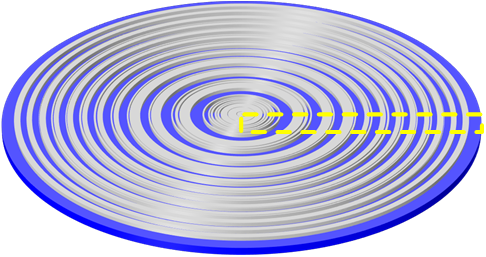}}
  \subfigure[NA=0.8]
  {\includegraphics[width=0.22\paperwidth]{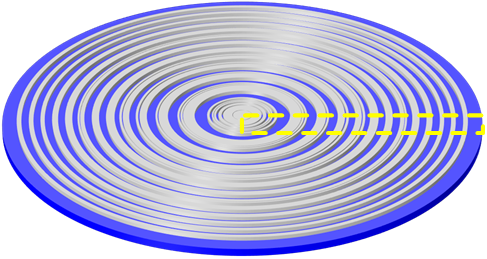}}
  \subfigure[NA=0.85]
  {\includegraphics[width=0.22\paperwidth]{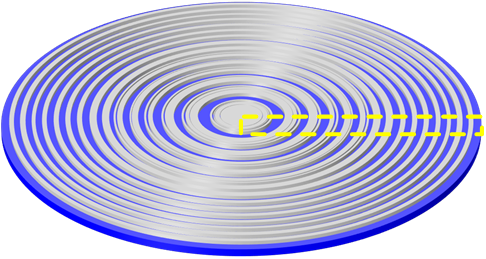}}
  \subfigure[NA=0.9]
  {\includegraphics[width=0.22\paperwidth]{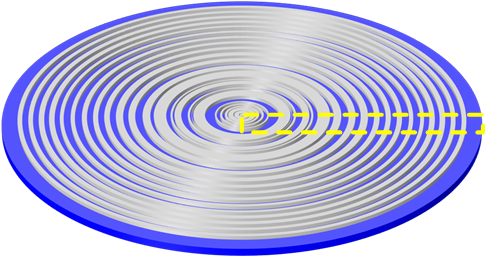}}\\
  \subfigure[Zoomed views of the regions marked by the dashed box in a$\sim$e.]
  {\includegraphics[height=0.25\paperwidth]{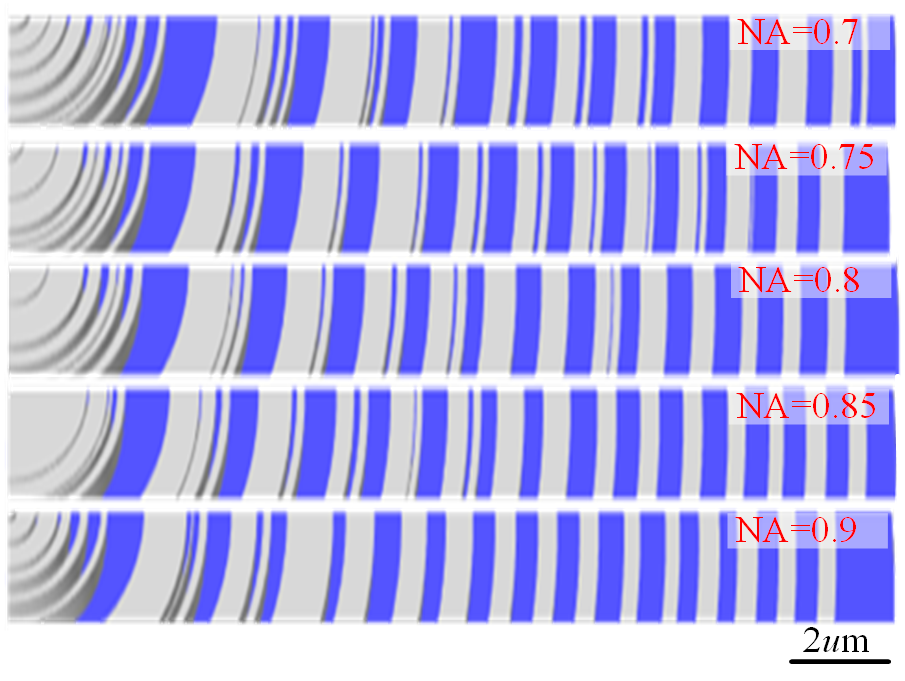}}
  \subfigure[Normalized electric field energy density in the focal plane.]
  {\includegraphics[height=0.28\paperwidth]{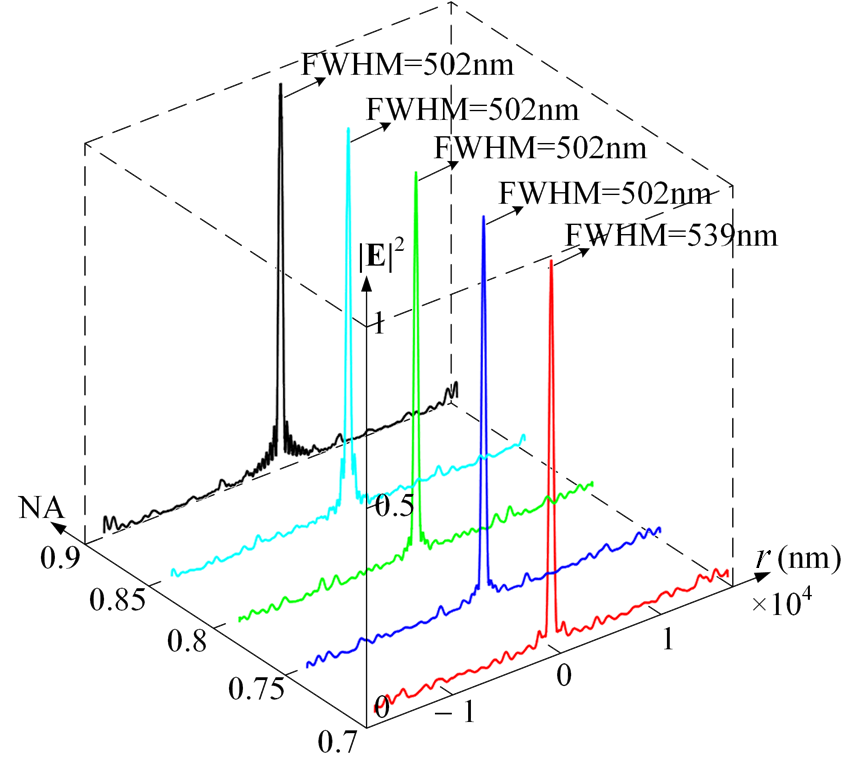}}
  \caption{Convex-like lenses with ring thickness 300~nm for an incident wavelength of 700~nm.}\label{FlatLenses300700}
\end{figure}

\begin{figure}[!htbp]
  \centering
  \subfigure[NA=0.7]
  {\includegraphics[width=0.22\paperwidth]{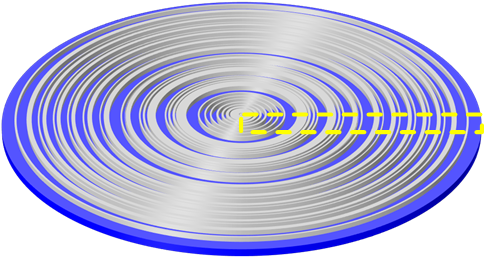}}
  \subfigure[NA=0.75]
  {\includegraphics[width=0.22\paperwidth]{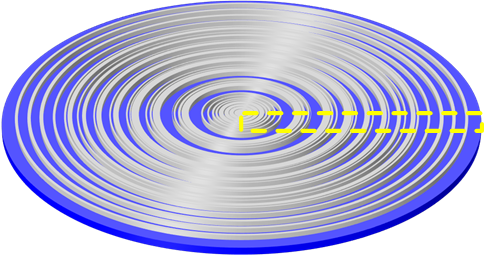}}
  \subfigure[NA=0.8]
  {\includegraphics[width=0.22\paperwidth]{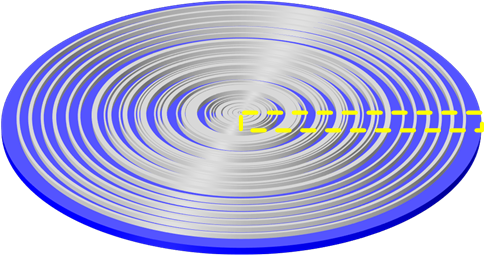}}
  \subfigure[NA=0.85]
  {\includegraphics[width=0.22\paperwidth]{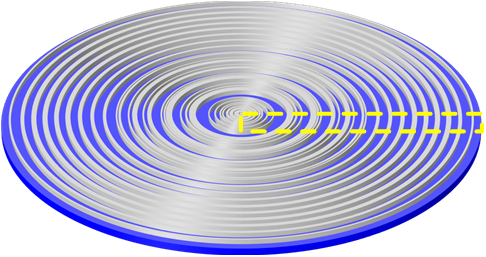}}
  \subfigure[NA=0.9]
  {\includegraphics[width=0.22\paperwidth]{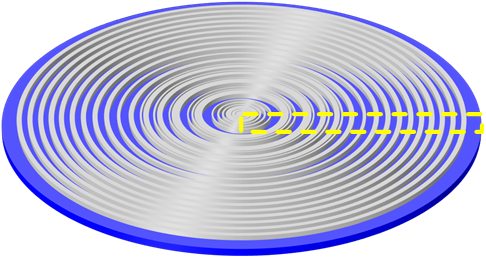}}\\
  \subfigure[Zoomed views of the regions marked by the dashed box in a$\sim$e.]
  {\includegraphics[height=0.25\paperwidth]{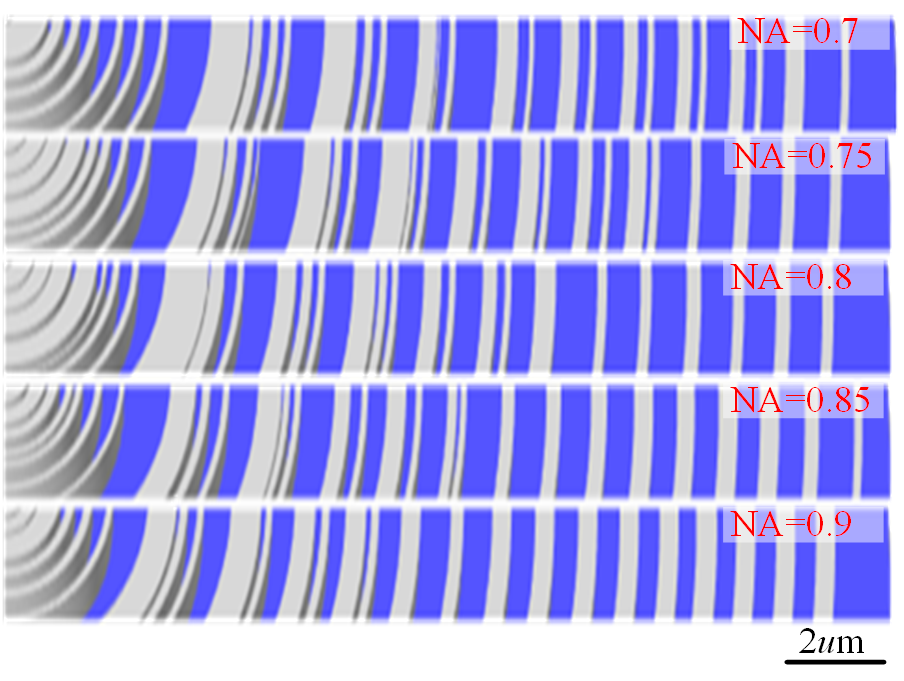}}
  \subfigure[Normalized electric field energy density in the focal plane.]
  {\includegraphics[height=0.28\paperwidth]{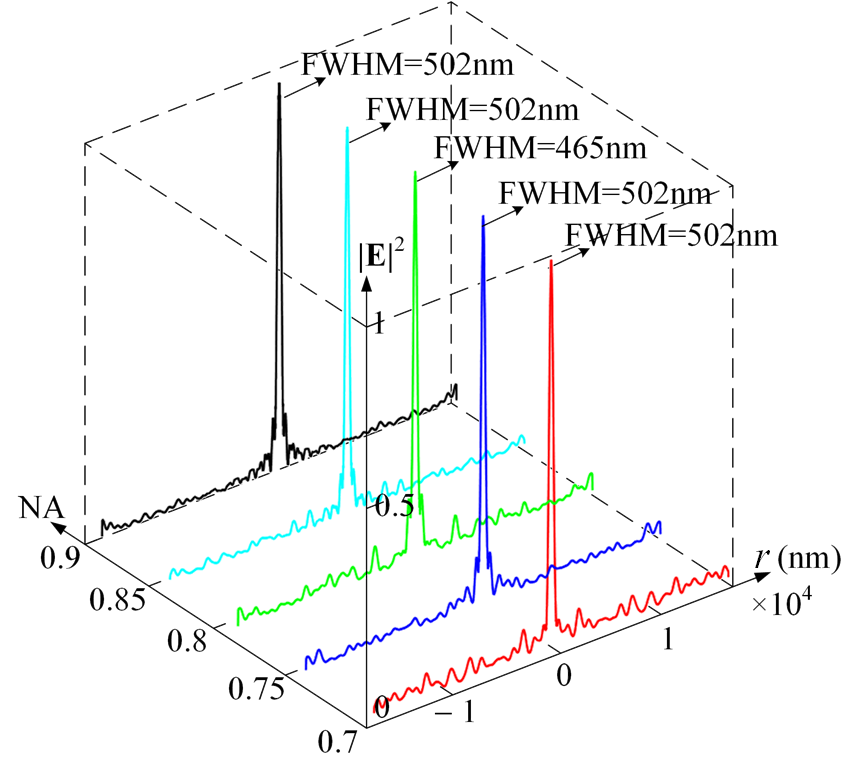}}
  \caption{Convex-like lenses with ring thickness 400~nm for an incident wavelength of 700~nm.}\label{FlatLenses400700}
\end{figure}

\begin{figure}[!htbp]
  \centering
  \subfigure[NA=0.7]
  {\includegraphics[width=0.22\paperwidth]{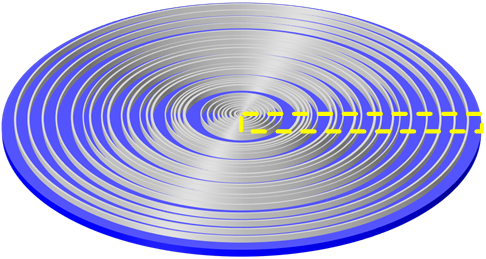}}
  \subfigure[NA=0.75]
  {\includegraphics[width=0.22\paperwidth]{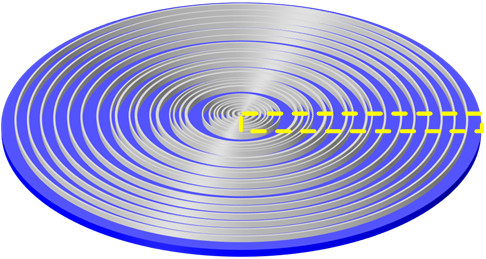}}
  \subfigure[NA=0.8]
  {\includegraphics[width=0.22\paperwidth]{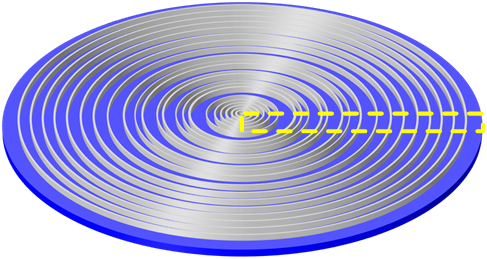}}
  \subfigure[NA=0.85]
  {\includegraphics[width=0.22\paperwidth]{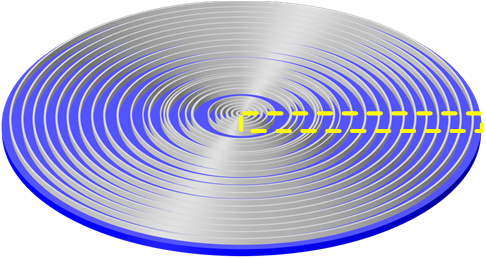}}
  \subfigure[NA=0.9]
  {\includegraphics[width=0.22\paperwidth]{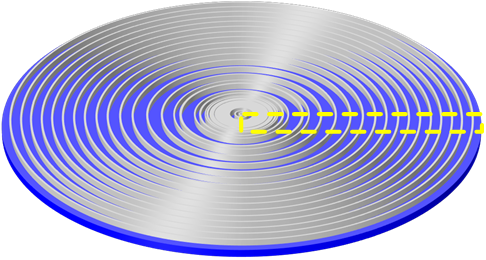}}\\
  \subfigure[Zoomed views of the regions marked by the dashed box in a$\sim$e.]
  {\includegraphics[height=0.25\paperwidth]{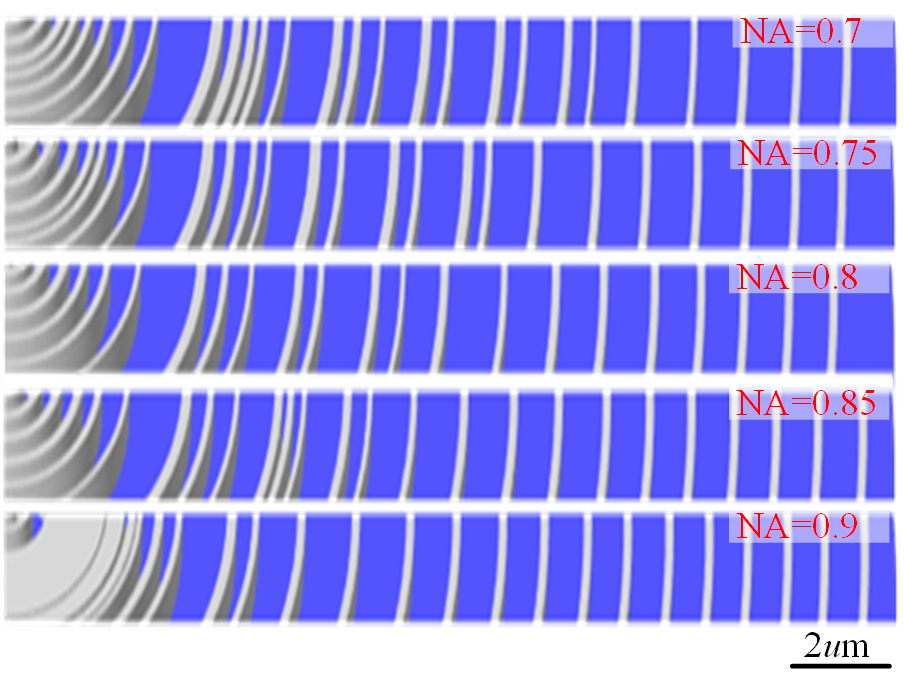}}
  \subfigure[Normalized electric field energy density in the focal plane.]
  {\includegraphics[height=0.28\paperwidth]{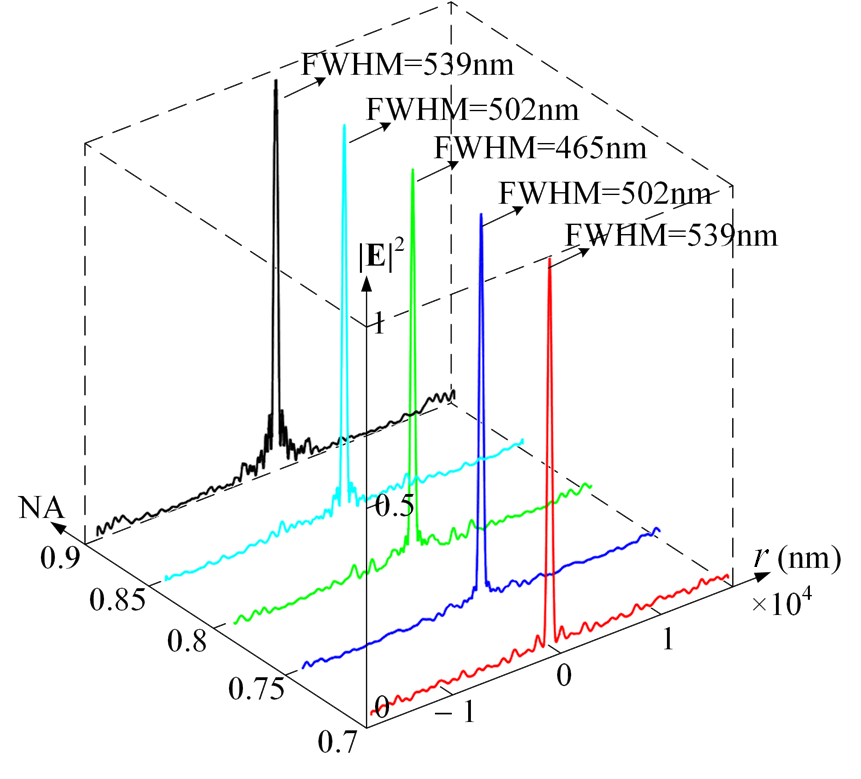}}
  \caption{Convex-like lenses with ring thickness 500~nm for an incident wavelength of 700~nm.}\label{FlatLenses500700}
\end{figure}

We now consider different convex lens designs. The rings are made of $\mathrm{TiO_2}$ and are formed on a quartz glass substrate, for which we use published permittivity values\cite{SiO2Para}. In the paper, for conciseness, we only show results for an incident wavelength of 700~nm, and the results for other wavelengths are placed in the supplementary documentation. For different metalens radii, we derive the designs in Figure \ref{Lamb700H200D} with the ring thickness 200~nm and numerical aperture 0.8, where the zoomed views of the rings are shown in Figure \ref{Lamb700H200D}f. By choosing the metalens radius to be 36$\mu$m, different designs are further derived in Figures \ref{FlatLenses200700}-\ref{FlatLenses500700} corresponding to variations in ring thickness and required numerical aperture, where the zoomed views of the rings are shown in Figures \ref{FlatLenses200700}f-\ref{FlatLenses500700}f. The normalized electric field energy density in the focal plane of each derived metalens (Figures \ref{Lamb700H200D}g-\ref{FlatLenses500700}g) shows that the full-width at half-maximum (FWHM) of the focal spot is around 500~nm, close to the diffraction limit ($\lambda/\left({2\mathrm{NA}}\right)$). In the derived metalenses, those with lower ring aspect ratios feature better manufacturability.

In the derived configurations, we observe that, as we increase the NA, the dielectric material gathers towards the center of a metalens. Thus more dielectric material appears in the central region of metalenses with higher NA. In this way, the interaction between the incident wave and center part of the metalens is strengthened, because a relatively stronger interaction is necessary to achieve a more drastic variation of the incident wave phase, and to achieve the necessary spherical wavefront, which ultimately converges to a focal spot that is in closer proximity to the metalens. It is conspicuous that secondary nanorings are prone to appear next to nanorings with relatively larger width. This characteristic is more pronounced in derived metalenses with smaller dielectric thickness. The presence of these secondary nanorings are prone to slow down the variation of the permittivity in the radial direction of the metalens, helping to ensure a phase discontinuity before and after the metalens, as required for the spherical wavefront.

\begin{figure}[!htbp]
  \centering
  \subfigure[Poynting vector distribution.]
  {\includegraphics[width=0.85\columnwidth]{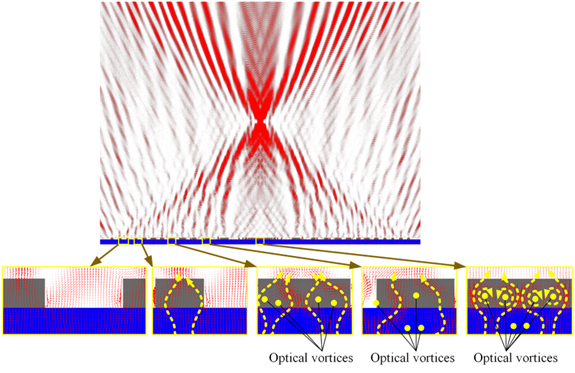}}
  \subfigure[Real part distribution of the field.]
  {\includegraphics[height=0.35\columnwidth]{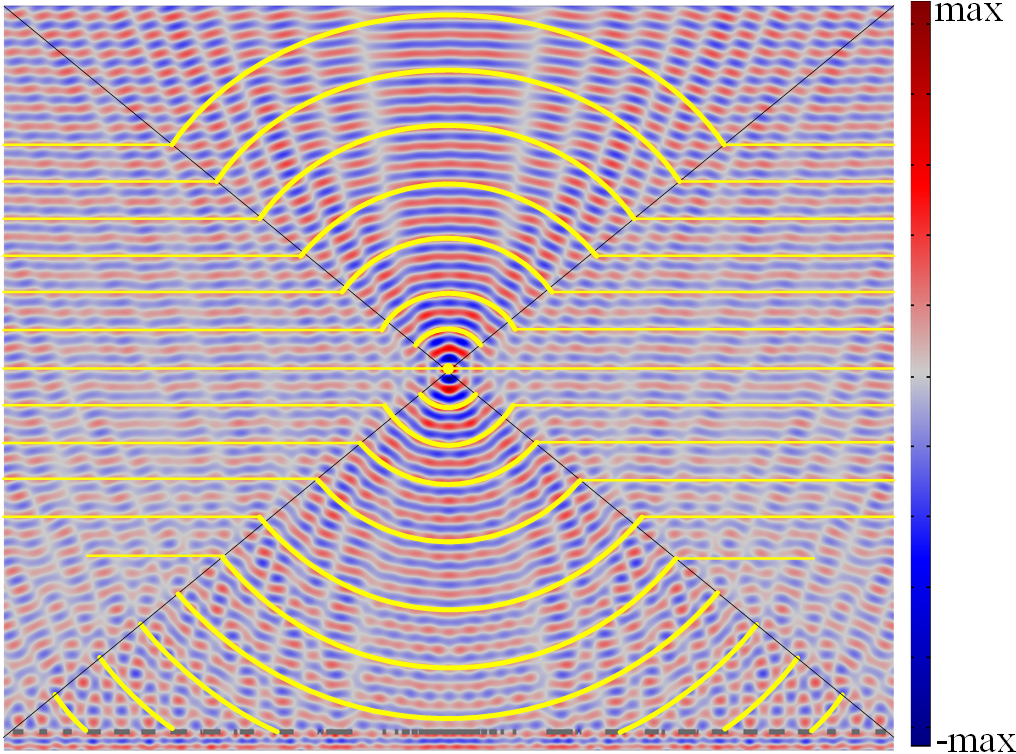}}
  \subfigure[Electric energy density distribution in the symmetrical plane.]
  {\includegraphics[height=0.35\columnwidth]{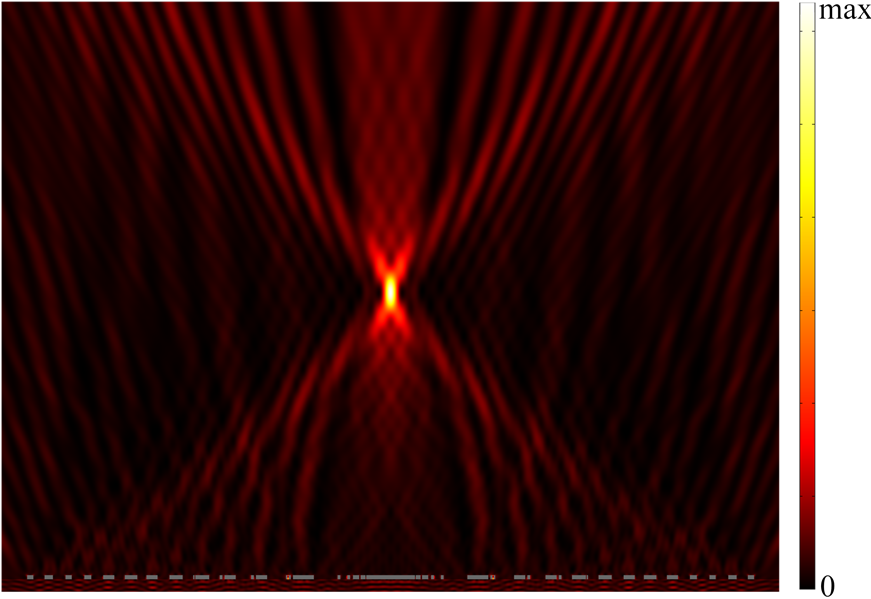}}
  \caption{(a) Poynting vector distribution in the metalens with incident wavelength 700~nm, ring thickness 200~nm, and $\mathrm{NA=0.8}$; (b) real part distribution of the field in the metalens, with spherical phase distribution indicated by the yellow-colored circular arcs; (c) electric energy density distribution in the symmetrical plane of the metalens.}\label{OpticalVortex20070008}
\end{figure}

The high-index TiO$_2$ material strongly interacts with the traversing light, so that the propagation speed $v$ is slowed down relative to the speed of light in free space ($v=c/n$). To achieve a focal spot at the desired position, energy vortexes are caused across the metalens (Figure \ref{OpticalVortex20070008}a). The energy vortices in the central part of the metalens are relatively strong, to equivalently prolong the propagation path of the photonic energy; whereas the vortexes are weaker in the outer part of the metalens. The energy vortices result in a phase match at the top of the metalens, and form a spherical phase distribution beyond the metalens (Figure \ref{OpticalVortex20070008}b), which is necessary to properly focus the energy in the manner of a convex-like lens (Figure \ref{OpticalVortex20070008}c). The Poynting vector distribution in Figure \ref{OpticalVortex20070008}a also shows that the impedance match between the substrate and the derived metalens is improved by the inversely designed TiO$_2$ structure. Reflections at the interface between the substrate and the metalens are strongly reduced, or equivalently, the effective transparency of the substrate is enhanced, so that more energy is passed and deposited at the focal spot.

\section{Extension}\label{Extension}

\begin{figure}[!htbp]
  \centering
  \subfigure[Axicon-like metalens with focal beam length equal to $0 \lambda$.]
  {\includegraphics[width=0.3\columnwidth]{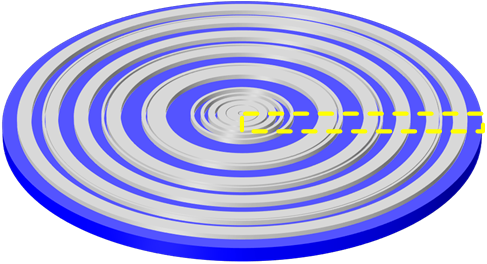}}
  \subfigure[Axicon-like metalens with focal beam length equal to $4 \lambda$.]
  {\includegraphics[width=0.3\columnwidth]{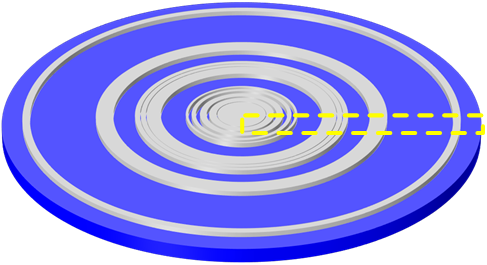}}
  \subfigure[Axicon-like metalens with focal beam length equal to $8 \lambda$.]
  {\includegraphics[width=0.3\columnwidth]{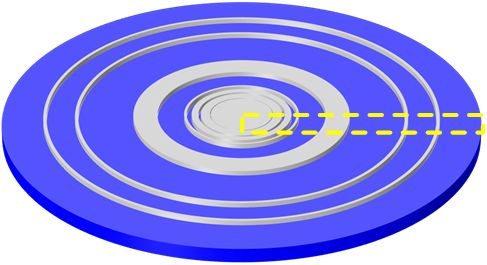}}\\
  \subfigure[Zoomed views of the regions marked by the dashed box in a$\sim$c.]
  {\includegraphics[height=0.25\paperwidth]{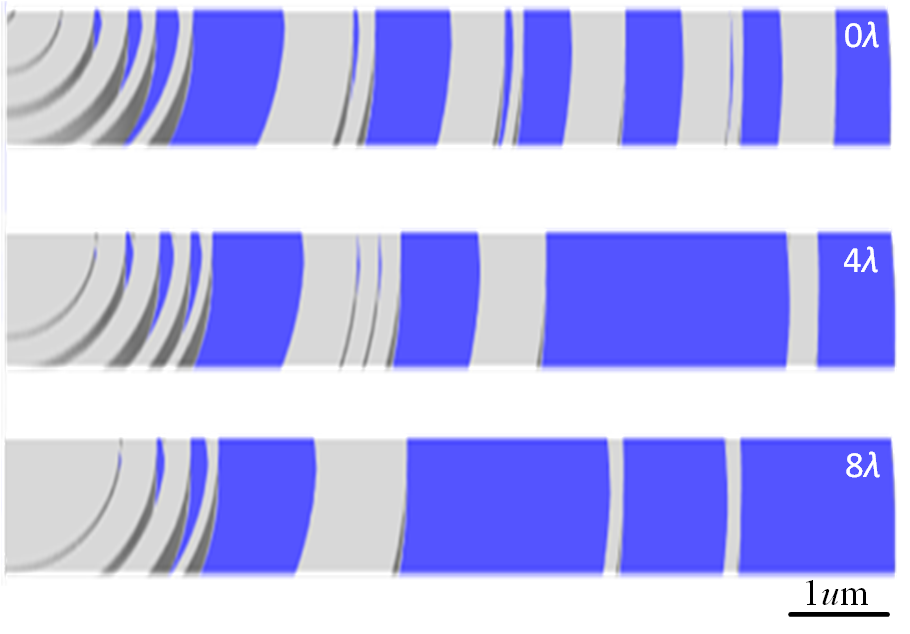}}
  \subfigure[Normalized electric field energy density in the focal plane localized at the center of the focal beam.]
  {\includegraphics[height=0.28\paperwidth]{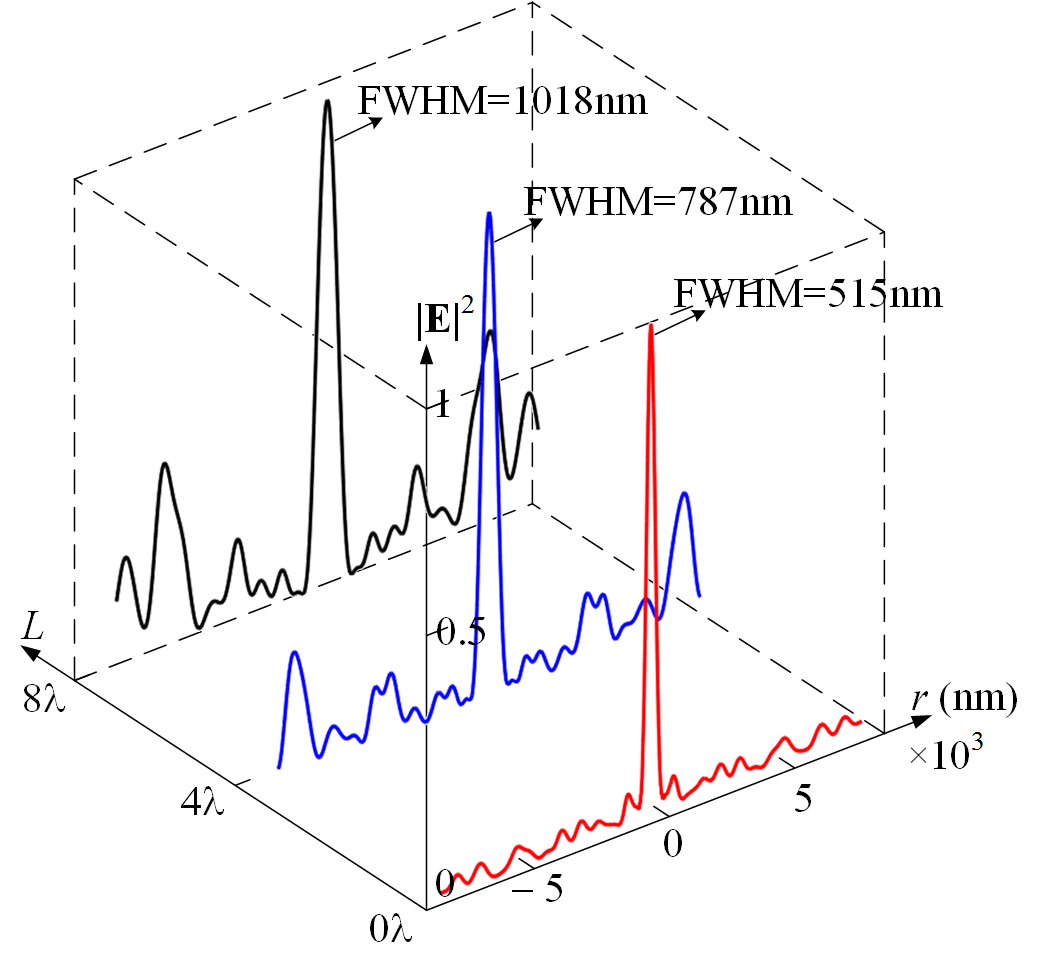}}
  \caption{Axicon-like metalenses with focal beam length set to an integer multiple of the incident wavelength.}\label{FlatAxicon}
\end{figure}

\begin{figure}[!htbp]
  \centering
  \subfigure[$b = 0\lambda$]
  {\includegraphics[width=0.2\paperwidth]{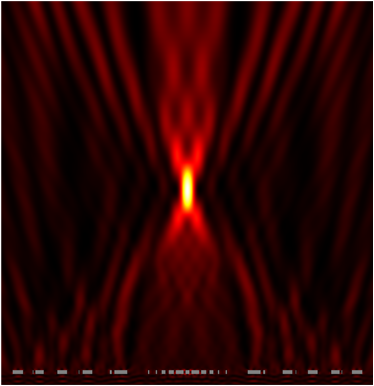}}
  \subfigure[$b = 4\lambda$]
  {\includegraphics[width=0.2\paperwidth]{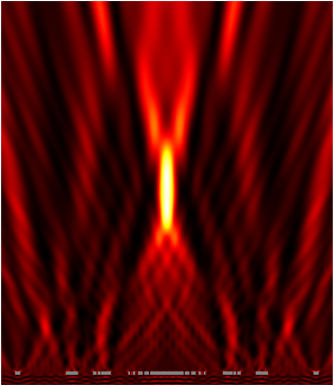}}
  \subfigure[$b = 8\lambda$]
  {\includegraphics[width=0.2\paperwidth]{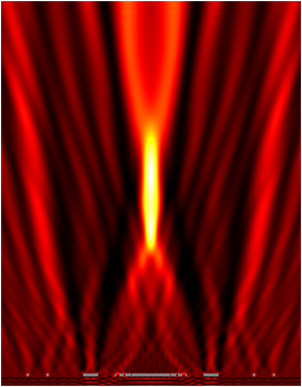}}\\
  \subfigure[$b = 0\lambda$, spherical wavefront.]
  {\includegraphics[width=0.2\paperwidth]{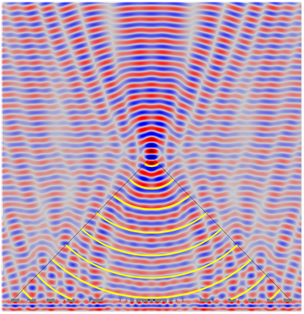}}
  \subfigure[$b = 4\lambda$, parabolic wavefront.]
  {\includegraphics[width=0.2\paperwidth]{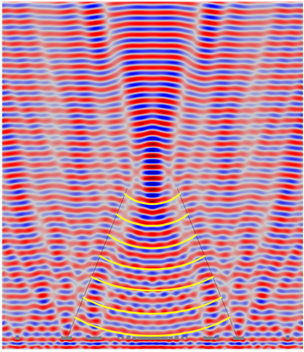}}
  \subfigure[$b = 8\lambda$, parabolic wavefront.]
  {\includegraphics[width=0.2\paperwidth]{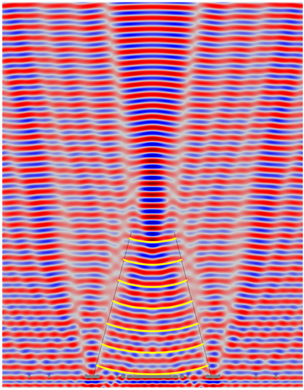}}
  \caption{(a-c) Electric field energy distribution in the axicon-like metalenses with focal beam $b$ respectively being 0, 4 and 8 fold of the incident wavelength $\lambda$; (d-f) real part magnetic field distribution in the axicon-like metalenses with focal beam respectively being 0, 4 and 8 fold of the incident wavelength, where the field phase of the wave after the metalenses are demonstrated as sketched by the yellow curves.}\label{AxiconField}
\end{figure}

The same methodology can be used to predict lenses with other focal distributions. For example, instead of a spot, an extended beam focus of axicon lenses can be targeted. Typically, axicons are conically shaped lenses that can focus light and create hollow Bessel beams \cite{McLeod1953,Manek1998,McLeod1960,Ren1990}. The inverse design objective for an axicon-like metalens is to maximize the minimal electric field density on the central line of the desired focal beam
\begin{equation}\label{FlatlensObj}
\begin{split}
J\left(\mathbf{E}\right) = \min_{\mathbf{p}_f\in I} \int_{\Omega_f} \left|\mathbf{E}\right|^2 \delta\left(\mathbf{p}-\mathbf{p}_f\right) \,\mathrm{d}v,
\end{split}
\end{equation}
where $\mathbf{p}_f\in\Omega_f$ is the point on the central line of the desired focal beam; $I$ is the central line segment of the desired focal beam. For an incident wavelength of 700~nm and a design domain radius of 9$\mu$m, axicon-like metalenses are derived in Figure \ref{FlatAxicon}a-c with zoomed views shown in \ref{FlatAxicon}d, where the NA of the metalens was set to 0.7 and the length of the desired focal beam was set to be a factor 0, 4 and 8 fold of the incident wavelength. The normalized electric field energy density in the focal plane localized at the center of the focal beam is plotted in Figure \ref{FlatAxicon}e. And the focusing efficiencies of the derived designs in Figure \ref{FlatAxicon}a-c are respectively $37.2\%$, $32.8\%$ and $26.6\%$, where longer focal beam corresponds to larger FWHM and lower focusing efficiency. The corresponding electric field energy distribution and real part magnetic field distributions are shown in Figure \ref{AxiconField}, where the field phase of the propagating wave after the metalenses are indicated by the black curves in Figure \ref{AxiconField}d-f. In Figure \ref{FlatAxicon}, the metalens with focal length equal to 0 fold of the incident wavelength, is identical to a convex-like metalens with a focal spot. From the field distributions of the inversely designed metalenses, one can conclude that the focusing efficiency of the derived metalenses decrease along with an increase in focal length, and the parabolic phase distributions are shaped in the zone after the metalenes to achieve a focal beam.

%

\section{Conclusions}\label{Conclusions}

This paper presented an inverse design method to predict efficient and compact metalenses with concentric-nanoring configurations. The procedure was implemented by a topology optimization approach. By focusing the electrical field energy at a specified position, a convex-like metalens was inversely designed with desired numerical aperture and a diffraction-limited focal spot, where the spherical wavefront behind the metalens was achieved based on a strong interaction between the concentric nanorings and the electromagnetic waves. The Poynting vector distribution in the derived convex-like metalens demonstrated the mechanism of the prediction, in which an optical vortex was generated in the nanorings to achieve a matching of the phase and impedance between the carrier substrate and free space and therefore enhance the transmission of the optical energy, and additionally to form a spherical wavefront which aids focusing. The inverse design method has been further extended to predict axicon-like metalenses with a strongly focused beam segment, in which a parabolic wavefront behind the metalens was formed.

These ultrathin optical structures have considerable advantages for numerous applications. For example, they would facilitate implementation in very tight spaces, as is often required in mobile applications. Because the target field distribution is conveniently customised, it is possible to use the method to predict unusual optical elements, such as focal spots with arbitrary geometrical pattern.

Although not proven yet, we are confident that the concentric-nanoring configuration of a derived metalens structure is manufacturable with available methods. For example, using a lift-off procedure \cite{Evans2015}, or dry etching \cite{Choi2013}, together with a nano-patterning approach such as electron-beam lithograpy, or two-photon nanolithography, it is feasible to pattern sputtered $\mathrm{TiO_2}$ structures on thin quartz glass substrates. It is our intention to investigate this as a next step.

\section{Acknowledgements}\label{Acknowledgements}

Y.D. was supported by the National Natural Science Foundation of China (No. 51405465), by a Guest Professor Fellowship at the Karlsruhe Institute of Technology (KIT) by the DFG Metacoil project (KO 1883/20-1), and by the National High Technology Program of China (No. 2015AA042604). J.G.K. is supported by an ERC Senior Grant (NMCEL, 290586) and by the Karlsruhe School of Optics and Photonics (KSOP). U.W. and J.G.K. are supported through the Cluster of Excellence BrainLinks-BrainTools (DFG, grant no. EXC 1086).


\begin{thebibliography}{}

\bibitem{Khorasaninejad2016}
M. Khorasaninejad, W. T. Chen, R. C. Devlin, J. Oh, A. Y. Zhu, F. Capasso, Metalenses at visible wavelengths: Diffraction-limited focusing and subwavelength resolution imaging, \textit{Science} 2016, \textbf{352}, 1190-1194.

\bibitem{Yu2014}
N. Yu, F. Capasso, Flat optics with designer metasurfaces, \textit{Nat. Mater.} 2014, \textbf{13}, 139-150.

\bibitem{Kildishev2013}
A. V. Kildishev, A. Boltasseva, V. M. Shalaev, Planar Photonics with Metasurfaces, \textit{Science} 2013, \textbf{339}, 1232009.

\bibitem{McLeod1953}
J. H. McLeod, The Axicon: A New Type of Optical Element, \textit{J. Opt. Soc. Am.} 1953, \textbf{44}, 592-597.

\bibitem{Hecht1997}
E. Hecht, \textit{Optics},  3rd ed., Addison  Wesley  Publishing  Company:  Reading,  MA,  1997.

\bibitem{Greegor2005}
R. B. Greegor, et al, Simulation and testing of a graded negative index of refraction lens, \textit{Applied Physics Letters} 2005, \textbf{87}, 091114.

\bibitem{Huang2008}
F. M. Huang, T. S. Kao, V. A. Fedotoc, Y. Chen, N. Zheludev, Nano-hole array as a lens, \textit{Nano Lett.}  2008,  \textbf{8},  2469-2472.

\bibitem{Huang2009}
F. M. Huang, N. Zheludev, Super-resolution without evanescent waves, \textit{Nano  Lett.} 2009,  \textbf{9},  1249-1254.

\bibitem{Rogers2012}
E. T. F. Rogers, J. Lindberg, T. Roy, S. Savo, J. E. Chad, M. R. Dennis,  N. I. Zheludev, A super-oscillatory lens optical microscope for subwavelength imaging, \textit{Nat. Mat.}  2012, \textbf{11}, 432-435.

\bibitem{Fattal2010}
D. Fattal, J. Li, Z. Peng, M. Fiorentino, R. G. Beausoleil, Flat dielectric grating reflectors with focusing abilities,  \textit{Nat. Phot.}  2010,  \textbf{4},  466-470.

\bibitem{Verslegers2009}
L. Verslegers, P. B. Catrysse, Z. Yu, J. S. White, E. S. Barnard, M. L. Brongersma, S. Fan, Planar Lenses Based on Nanoscale Slit Arrays in a Metallic Film, \textit{Nano Lett.} 2009, \textbf{9}, 235-238.

\bibitem{Pendry2000}
J. B. Pendry, Negative Refraction Makes a Perfect Lens, \textit{Phys.  Rev.  Lett.}  2000,  \textbf{85},  3966-3969.

\bibitem{Smith2004}
D. R. Smith, J. B. Pendry, M. C. K. Wiltshire; Metamaterials and negative refractive index, \textit{Science} 2004, \textbf{305},  788-792.

\bibitem{Liu2007}
Z. Liu, H. Lee, Y. Xiong, C. Sun, X. Zhang, Far-field optical hyperlens magnifying sub-diffraction-limited objects, \textit{Science}  2007,  \textbf{315},  1686.

\bibitem{Cai2010}
W. Cai, V. Shalaev, \textit{Optical  Metamaterials  Fundamentals  and  Applications}, 2010, Springer.

\bibitem{Hu2016}
J. Hu, C. H. Liu, X. Ren, L. J. Lauhon, T. W. Odom, Plasmonic Lattice Lenses for Multiwavelength Achromatic Focusing, \textit{ACS Nano} 2016, \textbf{10}, 10275-10282.

\bibitem{Boltasseva2008}
A. Boltasseva, V. M. Shalaev, Fabrication of optical negative-index metamaterials: recent advances and outlook, \textit{Metamaterials} 2008, \textbf{2}, 1-17.

\bibitem{Jin2014}
J. Jin, \textit{The finite element method in electromagnetics}, 2nd Edition, New York: John Wiley \& Sons, 2002.

\bibitem{Lazarov2011}
B. Lazarov, O. Sigmund, Filters in topology optimization based on Helmholtz type differential equations, \textit{Int. J. Numer. Methods Eng.} 2011, \textbf{86}, 765-781.

\bibitem{Wang2011}
F. Wang, B. S. Lazarov, O. Sigmund, On projection methods, convergence and robust formulations in topology optimization, \textit{Struct. Multidiscip. Optim.} 2011, \textbf{43}, 767-784.

\bibitem{Deng20161}
Y. Deng, J. G. Korvink, Topology optimization method for three dimensional electromagnetic waves, \textit{Proceedings of Royal Society A} 2016, \textbf{472}, 20150835.

\bibitem{Guest2004}
J. Guest, J. Prevost, T. Belytschko, Achieving minimum length scale in topology optimization using nodal design variables and projection functions, \textit{Int. J. Numer. Methods Eng.} 2004, \textbf{61}, 238-254.

\bibitem{Deng2015}
Y. Deng, Z. Liu, C. Song, J. Wu, Y. Liu, Y. Wu, Topology optimization based computational design methodology for surface plasmon polaritons, \textit{Plasmonics} 2015, \textbf{10}, 569-583

\bibitem{Hinze2009}
M. Hinze, R. Pinnau, M, Ulbrich, S. Ulbrich, \textit{Optimization with PDE constraints}, Springer, Berlin, 2009.

\bibitem{comsol}
http://www.comsol.com

\bibitem{Svanberg1987}
K. Svanberg, The method of moving asymptotes: a new method for structural optimization. \textit{Int. J. Numer. Meth. Engng.} 1987, \textbf{24}, 359-373.

\bibitem{SiO2Para}
http://refractiveindex.info


\bibitem{Manek1998}
I. Manek, B. Y. Ovchinnikov, R. Grimm; Generation of a hollow laser beam for atom trapping using an axicon, \textit{Opt. Commun.} 1998, \textbf{147}, 67-70.

\bibitem{McLeod1960}
J. H. McLeod; Axicons and their uses, \textit{J. Opt. Soc. Am.} 1960, \textbf{50}, 166-169.

\bibitem{Ren1990}
Q. Ren, R. Birngruber; Axicon: A new laser beam delivery system for corneal surgery. \textit{IEEE J. Quantum Electron.} 1990, \textbf{26}, 2305-2308.

\bibitem{Evans2015}
 C. C. Evans, C. Liu, J. Suntivich; J. Low-loss titanium dioxide waveguides and resonators using a dielectric lift-off fabrication process. \textit{Optics Express} 2015, \textbf{23}, 11160.

\bibitem{Choi2013}
K. R. Choi, J. C. Woo, Y. H. Joo, Y. S. Chun, C. I. Kim, Dry etching properties of TiO. \textit{Vaccum} 2013, \textbf{92}, 85-89.


\end{thebibliography}
\end{document}